\documentclass[]{article}
\usepackage{graphicx}                    % For eps figures, newer & more powerfull
\usepackage[square]{natbib}
\begin{document}
\title{Ten Years of the Solar Radiospectrograph ARTEMIS--IV}
\author{	C. Caroubalos \\University of Athens, 15783 Athens, Greece \\
		C. E. Alissandrakis\\University of Ioannina, 45110 Ioannina, Greece\\
		A. Hillaris, P. Preka--Papadema\\University of Athens, 15783 Athens, Greece\\
		P. Tsitsipis, A. Kontogeorgos, V. Petoussis\\ Technological Education Institute of Lamia, Lamia, Greece\\
		C. Bouratzis\\University of Athens, 15783 Athens, Greece
		J.-L. Bougeret, G. Dumas\\Observatoire de Paris, CNRS UA 264, 92195 Meudon Cedex, France\\
		A. Nindos\\University of Ioannina, 45110 Ioannina, Greece}
\maketitle

\begin{abstract}
The Solar Radiospectrograph of the University of Athens 
(ARTEMIS-IV\footnote{Appareil de Routine pour le Traitement et l' Enregistrement Magnetique de l' Information Spectral }) 
is in operation at the Thermopylae Satellite Communication
Station since 1996. The observations extend from the base of the Solar Corona (650 MHz) to about 2 
Solar Radii (20 MHz) with time resolution 1/10-1/100 sec. 
The instruments recordings, being in the form of dynamic spectra, measure radio flux as a function of height in 
the corona; our observations are combined with spatial data from the Nancay Radioheliograph whenever the need 
for 3D positional information arises. 
The ARTEMIS-IV contribution in the study of solar radio bursts is two fold-- Firstly, in investigating new spectral characteristics since its high sampling rate facilitates the study of fine structures in radio events. On the other hand 
it is used in studying the association of solar bursts with interplanetary phenomena because of its extended 
frequency range which is, furthermore, complementary to the range of the WIND/WAVES receivers and the observations 
may be readily combined. This reports serves as a brief account of this operation. Joint observations with 
STEREO/WAVES and LOFAR low frequency receivers are envisaged in the future.
\end{abstract}

\section{Introduction}

The radio emission of solar flares and transients includes a plethora of emission processes which
serve as diagnostic tools 
for the analysis of non-thermal electron distributions, plasma waves and turbulence (cf. \cite{Benz} for a review). 
They are caused by plasma instabilities driving various wave modes that in turn may emit observable 
radiation. Some of the processes are reasonably well understood others are open research issues. 
Solar radio spectrographs have revealed a large variety of Solar radio--bursts taking
place in the Solar Corona. As observations at different wavelengths sample different heights and
physical conditions in the solar atmosphere, with longer wavelengths referring to higher 
heights above the photosphere, these phenomena can in principle be probed from the bottom of the corona to 
large distances in the interplanetary medium. This report is a brief account of a decade of the ARTEMIS--IV 
Solar Radiospectrograph operation in the field of observational analysis of solar radio bursts.

\section{Operation and System Architecture of the ARTEMIS--IV Solar Radiospectrograph}
The Solar Radiospectrograph of the University of Athens (ARTEMIS-IV) is in operation at the
Thermopylae Satellite Communication Station since 1996. The observations extend from the base 
of the Solar Corona (650 MHz) to about 2 Solar Radii (20 MHz) It offers high quality recordings
with time resolution 1/10-1/100 sec, in the form of dynamic spectra; which is radio flux as a
function of height in the corona.
%-------------------------------------------------------------------------------------
\begin{figure}
\centerline{\includegraphics[width=\textwidth]{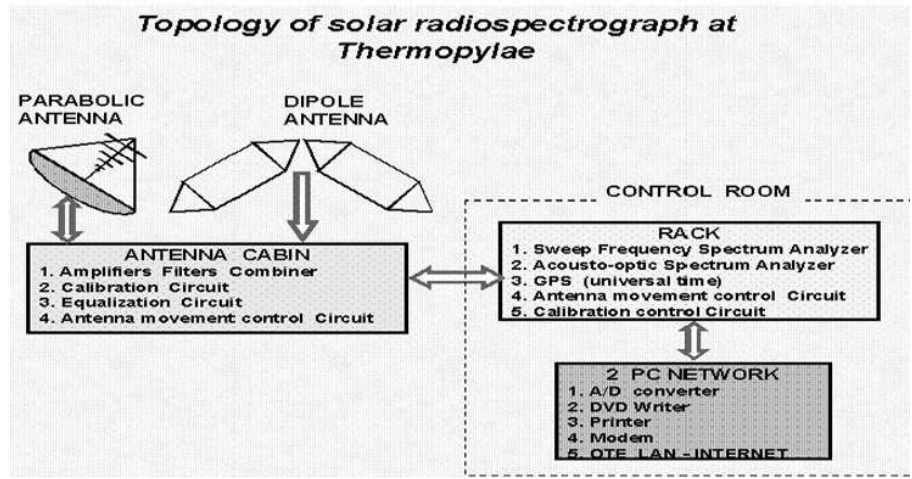}}
\caption{Block Diagram of the The Artemis--IV radiospectrograph architecture}
\label{CONFIGURATION}
\end{figure}
%-------------------------------------------------------------------------------------
\begin{figure}
\begin{minipage}[t]{6cm}
\begin{center}
\includegraphics[width=6cm]{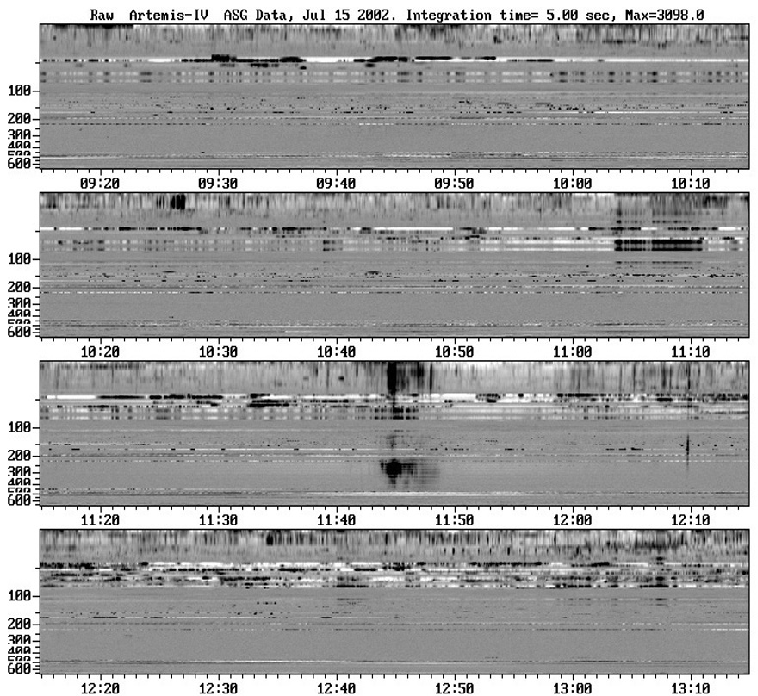}
\end{center}
\end{minipage}
\hfill
\begin{minipage}[t]{6cm}
\begin{center}
\includegraphics[width=6cm]{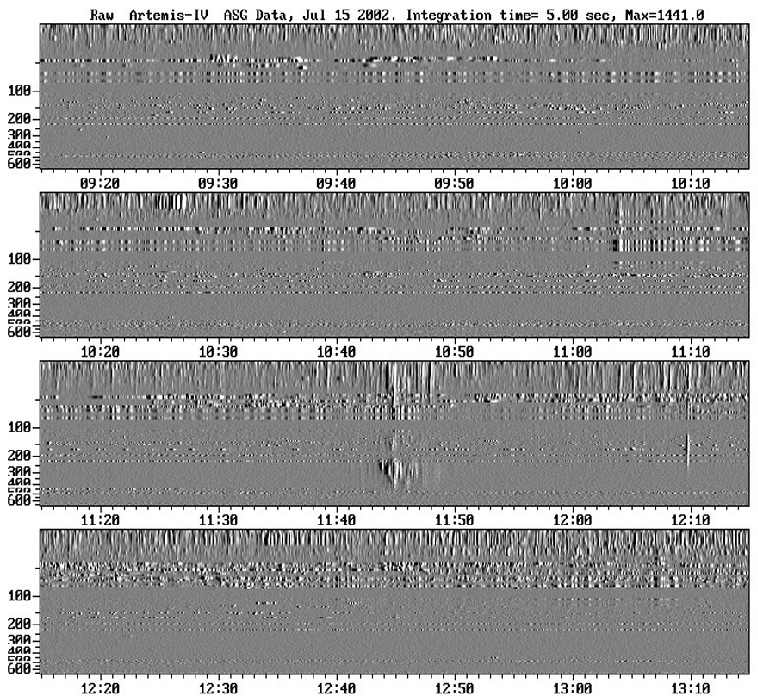}
\end{center}
\end{minipage}
\caption{An ASG QUICKLOOK Intensity Spectrum July 15 2001. The parallel lines are terrestrial interference. 
	 LEFT: Intensity Spectrum, RIGHT: Differential Spectrum.}
\label{QUICKLOOK01}
\end{figure} 
%-------------------------------------------------------------------------------------
\begin{figure}
\begin{minipage}[t]{6cm}
\begin{center}
\includegraphics[width=6cm]{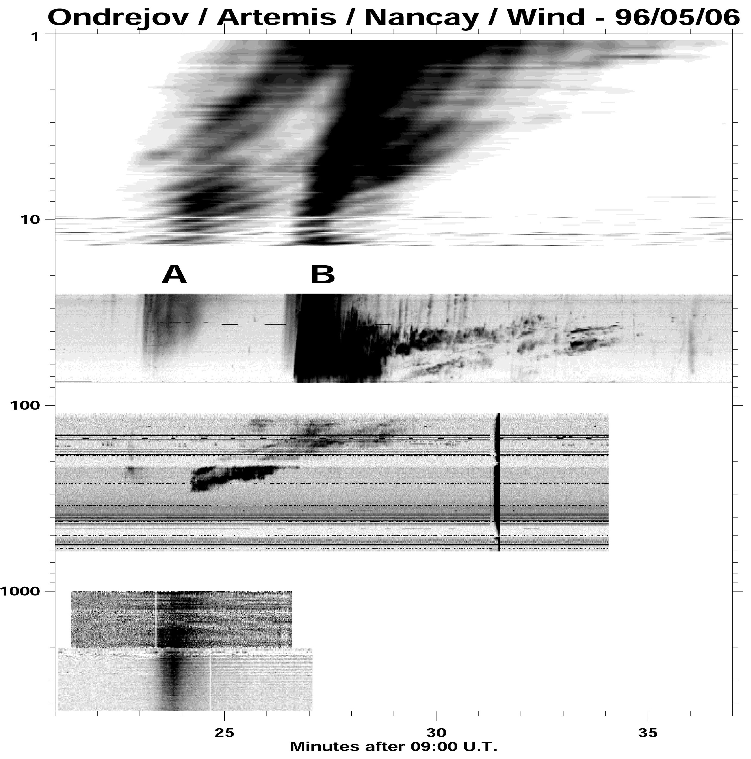}
\end{center}
\end{minipage}
\hfill
\begin{minipage}[t]{6cm}
\begin{center}
\includegraphics[width=6cm]{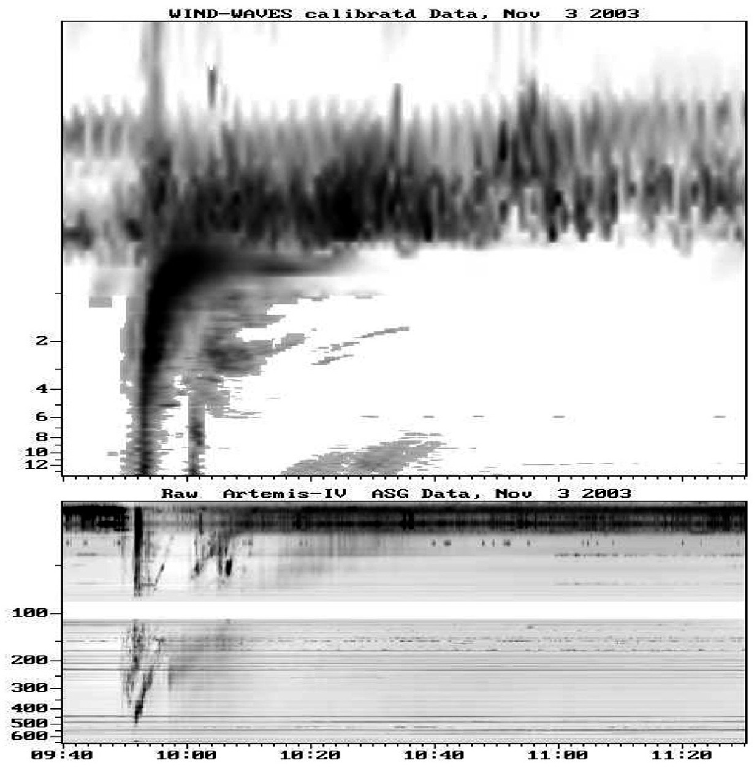}
\end{center}
\end{minipage}
\caption{Global Dynamic Spectra. LEFT: The May 6, 1996 event from 4.2GHz to 1
MHz.  From Bottom to Top: Ondrejov Radiospectrogram; ARTEMIS-IV data;
Nan\c cay Decametric Array; Wind/Waves/RAD2 (\cite{Bougeret}) RIGHT: 
Combined ARTEMIS--IV and WIND--WAVES spectra of the November 3, 2003 event.}
\label{MAY96}
\end{figure} 
%-------------------------------------------------------------------------------------
\begin{figure}
\centerline{\includegraphics[width=\textwidth]{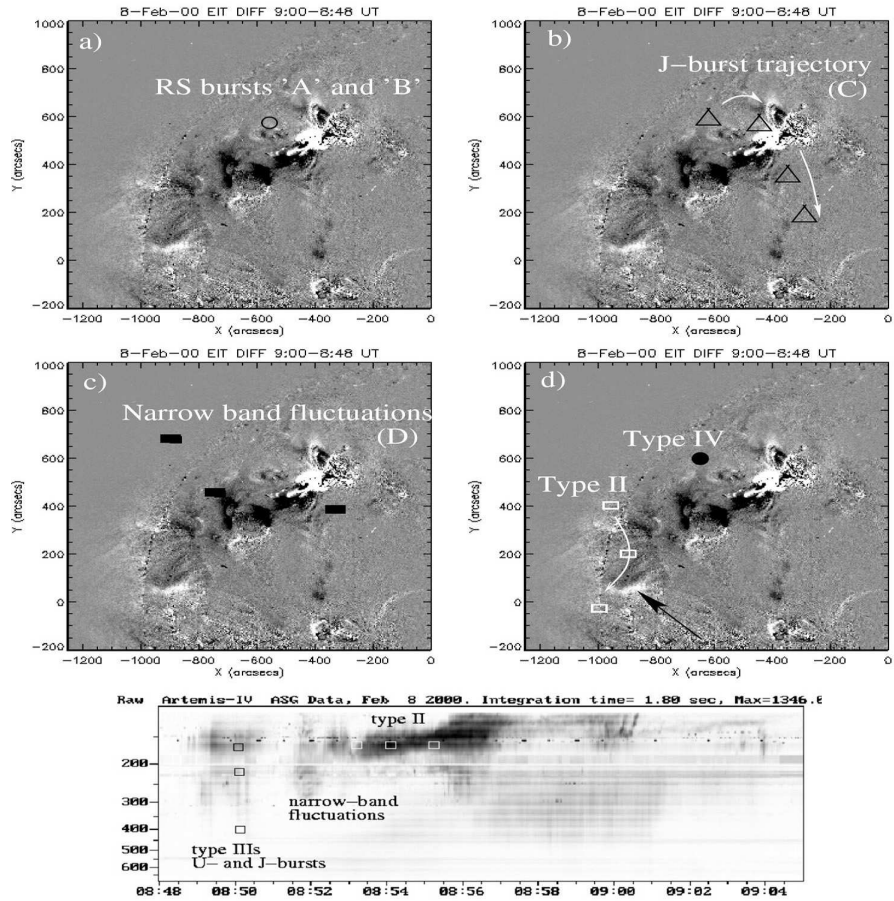}}
\caption{The February 08, 2000 Event. 
UPPER FOUR PANELS: NRH Positions of reverse slope (RS) and J-bursts and narrowband
fluctuations overlayed on an EIT difference image.
BOTTOM PANEL: Artemis--IV Dynamic radio spectrum, 
showing J-bursts, narrowband fluctuations,and the 2nd harmonic lane of the type II burst. 
Black and white circles indicate the radio sources that are imaged.} 
\label{}
\end{figure}
%-------------------------------------------------------------------------------------
\begin{figure}
\centerline{\includegraphics[width=\textwidth]{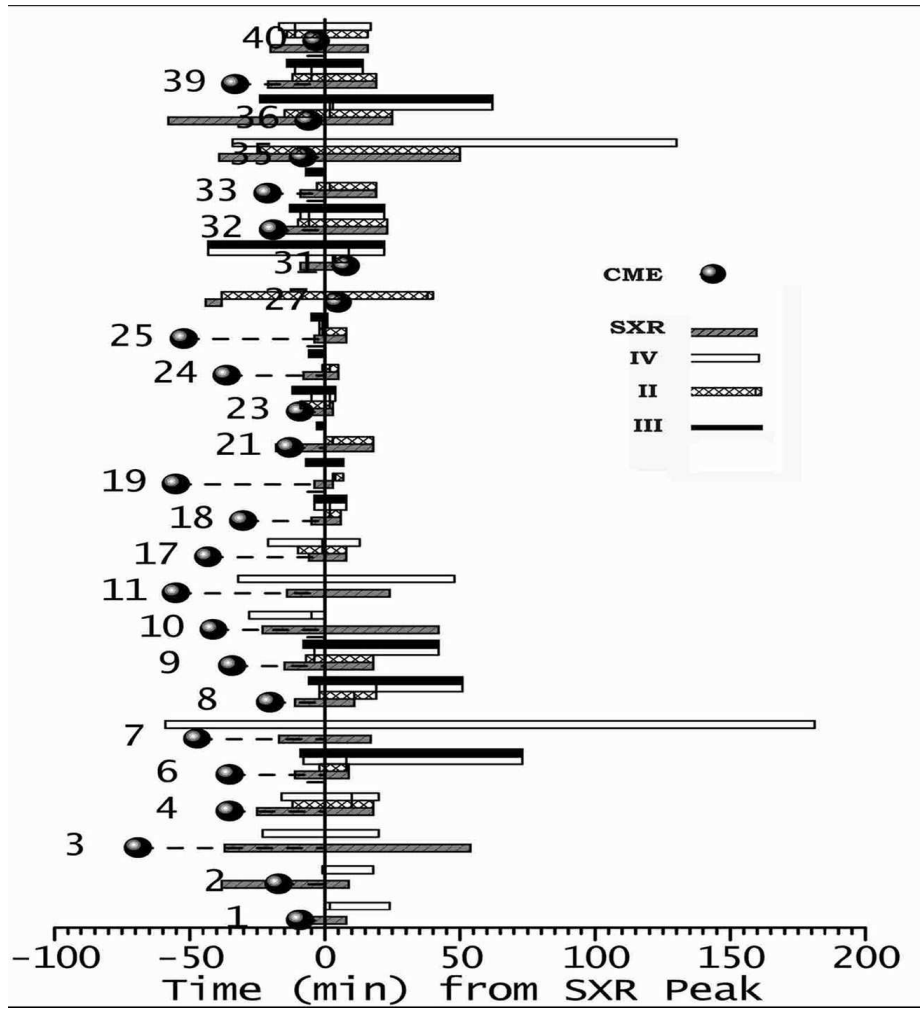}}
\caption{Schematic representation of the total activity of the CME associated
events. It depicts durations of SXR flares, Type III, II, IV activity
and extrapolated CME launch time; all times are measured from
the SXR flare peak~\cite{Caroubalos04}.} 
\label{CMEShockFlare}
\end{figure}
%-------------------------------------------------------------------------------------
\begin{figure}
\centerline{\includegraphics[width=\textwidth]{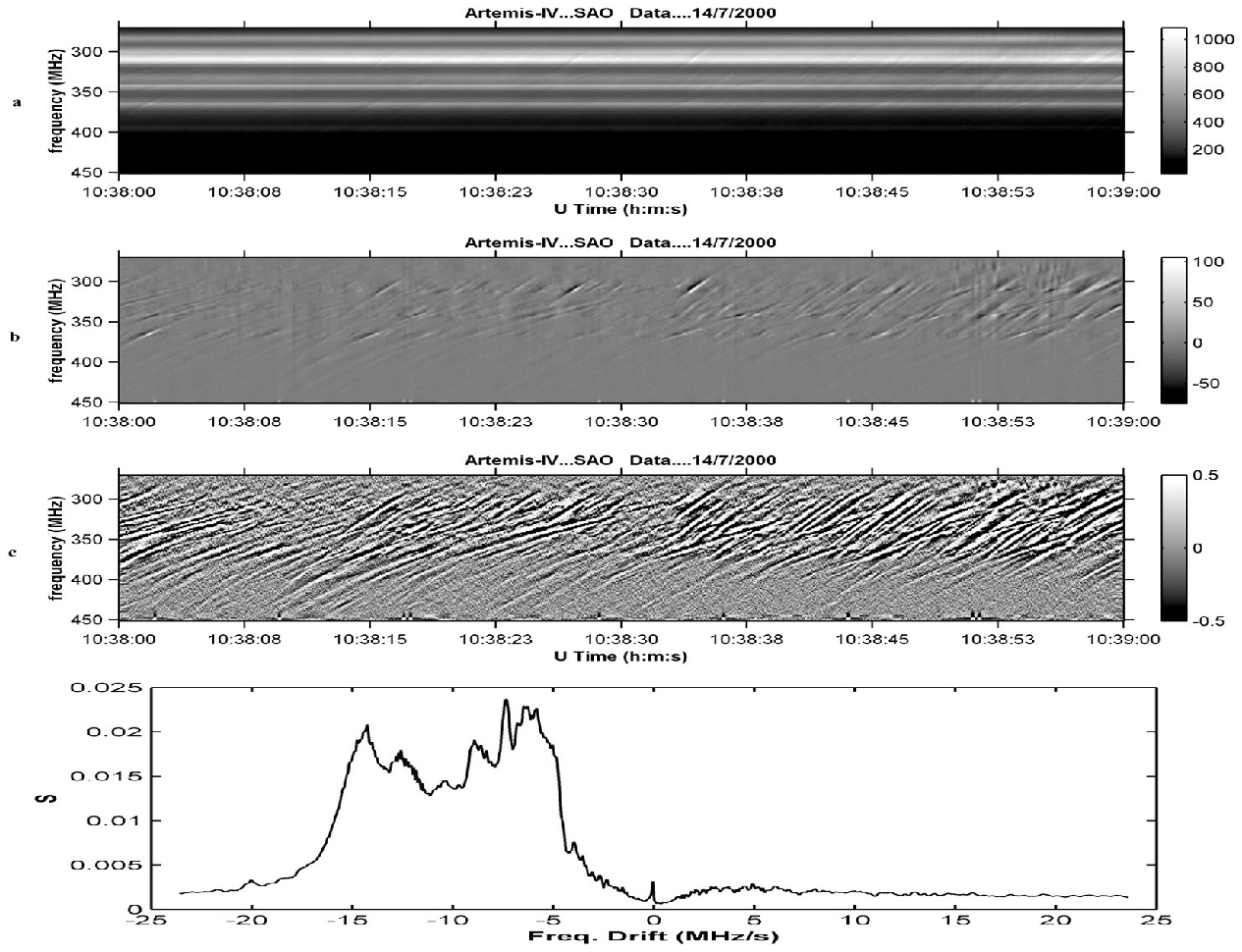}}
\caption{Detection of fiber bursts embedded in a type IV continuum
on a dynamic spectrum recorded by ARTEMIS-IV on the 14 July 2000 \cite{Caroubalos01b}. 
Top: Dynamic Spectrum of a solar continuum; the lines parallel to the 
time axis represent fixed frequency terrestrial interference. 
Second panel from the top: Same images after a high pass filtering along the time axis; 
this has eliminated the interference and has suppressed the continuum enhancing the embedded fine structure. 
The latter consists of pulsations \& fibers.Third panel from top: 
High pass filtering along the frequency axis has suppressed the pulsations
enhancing the {\em fibers\/}). 
Bottom: The application of the {\emph{Slope Detection Algorithm}},~\cite{Tsitsipis},
provides a distribution of fiber {\emph{Slopes}} for the filtered dynamic spectrum;
these correspond to the distribution of frequency drift rates for the fibers.} 
\label{FE}
\end{figure}
%-------------------------------------------------------------------------------------

The radio emission is collected 
by two antennas. Firstly, a 7-meter parabolic antenna  
with equatorial mounting covering the frequency range 100--650 MHz; it tracks 
the Sun in Declination and Hour Angle. Secondly, a stationary {\emph {inverted V}} antenna is used for the 
20 to 110 MHz range. The incoming signal from the two antennas is combined and 
then shared between two receivers (Figure \ref{CONFIGURATION}). 
The sweep frequency receiver (Analyseur de Spectre Globalor ASG), 
covers the range of 20-650 MHz at 10 samples/sec, with a dynamic range of 70 db. 
The Acousto-Optical receiver (Spectrographe Acousto Optique or SAO), which covers the range of 270-450 MHz 
at 100 samples/sec, with a dynamic range of 25 db. A detailed description of the receivers can be found in 
\cite{Caroubalos01a}. 

The data acquisition is performed by two PCs (equipped with 12 bit, 225 ksamples/sec 
DAC, one for every receiver), connected through Ethernet. The daily operation is fully 
automated.  

The recordings are in the form of dynamic spectra from the ASG and the SAO. 
The ASG gives good observations of strong, 
broad band events, such as type III groups and type II/type IV bursts. The SAO, on the other hand, has a 
higher sensitivity and lower noise compared to classic sweep--frequency receivers, although its dynamic 
range is rather limited; there is a wealth of fine structure comprising of weak, small time scale 
as well as narrow band features observable at this resolution.

In order to enhance small fluctuations and features which are usually undetected as they are
superimposed on high level activity the time derivative of
the intensity for each frequency channel is often used. Positive derivatives are coded in black,
negative in white and all intermediate grey levels have been adjusted to obtain an
optimal presentation of the events. 

The daily spectra of the ASG with a reduced time resolution of 5 seconds (Quick Looks)
are available on line from the ARTEMIS-IV web page\footnote{http://www.cc.uoa.gr/artemis} (cf figure
\ref{QUICKLOOK01} for an example). Higher resolution data are available on request. 

\section{Observational Results}

In the following subsections, we review in brief, observational results of solar radio bursts, recorded
by ARTEMIS-IV and, mostly combined with data from other sources:

\subsection{Joint Observations from the base of the Solar Corona to Earths Orbit: The SA Electrons}

The SA Electrons constitute a class of intense interplanetary type III bursts closely
associated to meter-wave (coronal) type II bursts. The acronym SA is interpreted either 
as {\emph{Shock Accelerated}} or as {\emph{Shock Associated}}; thus expressing an uncertainty on the origin 
of the electrons responsible for the SA events. They are thought to be  produced by electrons accelerated
by shock waves in the solar corona as an extension of {\emph{herringbone structures}}
or by electrons accelerated in the chromosphere in the same energy release that has launched the shock.

An almost complete frequency coverage of an SA radio
event and related phenomena observed on May 6, 1996 at 9:27 UT 
was presented for the first time (\cite{Bougeret}, cf. also Figure \ref{MAY96}). 
It was observed from the base of the solar corona up to almost 1 AU 
from the Sun by the following radio astronomical instruments: 
\begin{itemize}
\item{the Ondrejov spectrometer operating in 4.5--1 GHz (radiation produced near the chromosphere)}
\item{The Artemis-IV spectrograph operating in 600--110MHz (1.1--1.4R  from sun center)}
\item{The Nan\c cay Decametric Array Decameter Array in 75--25 MHz (1.4--2 R)}
\item{The RAD2 and RAD1 radio receivers on the WIND spacecraft covering the range from 14 MHz--20 kHz (3R--1AU). }
\end{itemize}
The observations show that the SA event starts from a coronal type II radio burst which traces the 
progression of a shock wave through the corona above {$1.8R-2R$} from the sun center. 
This SA event had no associated radio emission in the decimetric-metric range, thus there is no evidence for 
electron injection in the low/middle corona and, as a case study, corroborates the Shock Acceleration.

\subsection{CME Events Associated with Large Scale Field Line Openings and Disappearing Transequatorial Loops}

Disappearing transequatorial loops have been observed on several occasions in association with the onset of 
coronal  mass ejections. The opening of the CME magnetic field can be revealed by radio observations since the radio emission sources trace the propagation paths of electrons (along the loops before they open up completely). 

A multi--wavelength analysis of two events (February 8, 2000, cf. \cite{Pohjolainen05} 
\& May 2, 1998 cf. \cite{Pohjolainen01}) was performed in order to test the transequatorial loop 
disappearance association with the CME lift off. Observational data were obtained from:

\begin{itemize}
 \item{SOHO LASCO,~in white light).}
\item{SOHO EIT,~in EUV}
\item{YOHKOH,~in SXR}
\item{The ARTEMIS--IV}
\item{The Nan\c cay Radioheliograph (NRH).}
\end{itemize}

This study provided combined observations of large--scale loop--like SXR \& EUV dimmings 
coincident, on disk, with type III (J) family bursts tracing the abovementioned loops and type IV emission 
marking the CME lift--off. This evidence in support of large scale field line opening preceding the lift--off 
of material corroborates the {\emph{breakout}} model of CME launch formulated by \cite{Antiochos}.

\subsection{CME--Flare Events Associated with Coronal Waves}

The question whether a given coronal wave was driven by a CME or it was ignited by a flare is an open~
research issue; this issue was addressed in a {\emph{case study}}(\cite{Vrsnak}) of the 
large flare/CME event that occurred close to the west solar limb on 3 November 2003. The comprehensive 
multi--wavelength coverage of this event, from radio to HXR, marked this event as an appropriate 
candidate for this type of analysis. The data included:
\begin{itemize}
\item{Ha Filtergrams from the Kanzelhohe Solar Observatory (KSO) }
\item{X--ray Images in SXR  from GOES-12 and in HXR from the Reuven Ramaty High Energy Solar Spectroscopic Imager}
\item{EUV images from SOHO EIT}
\item{Dynamic spectra from: The ARTEMIS--IV, 
the Radiospectrograph of the Solar Radio Laboratory of Izmiran, 
the Potsdam--Tremsdorf Radiospectrograph, the RT4/5 spectrograph of Ondrejov, 
the PHOENIX-spectroghraph of ETH Zurich \& the RAD1 radio receiver on the WIND spacecraft.}
\item{The Nan\c cay RADIOHELIOGRAPH.}
\item{Magnetograms from the Michelson Doppler Imager onboard SoHO}
\item{White light observations from SOHO LASCO.}
\end{itemize}
The analysis indicated that the physical relationship between CME/flare and the coronal wave (shock) may be quite 
complex. In this case it was found that the coronal shock that produced the Moreton wave and the associated 
type II burst was generated by the energy release in the flare rather than by the CME expansion; 
this evidence in support of the flare initiated shock however does not preclude the CME initiation scenario.

\subsection{Association of type II and type IV Burst with other type of Solar Active Phenomena}

A catalogue of the type II and type IV solar radio bursts in the 110--687 MHz range, observed with ARTEMIS-IV 
in 1998--2000 was presented (\cite{Caroubalos04}); these were compared with the LASCO archives of Coronal Mass 
Ejections and the Solar Geophysical Reports of solar flares (Ha \& SXR) and examined for possible associations. 
The main results of this work corroborating previous works (\cite{Andrews},~\cite{Classen}, \cite{Zhang}, 
\cite{Sheeley}) are:
\begin{itemize}
 \item{68\% of the catalogue events were associated with CMEs.}
\item{67\% of the type II events were associated with CMEs, in accordance with previous results. 
T	his percentage rises to 79\% in the case of composite type II/IV events.}
\item{77\% of the type IV continua were associated with CMEs, 
	which is higher that the CME--type II association probability.}
\item{The type II associated CMEs had an average velocity of $(835\pm380) km/s$, 
	while the CMEs without type IIs had an average velocity of $(500\pm150) km/s$.}
\item{All events, but one, were well associated with Ha and/or SXR flares.}
\item{Most of the CME launch times precede by 5--60 min (30 min on average) the associated SXR flare peak; 
	an important fraction (72\%) precede the flare onset as well.}
\item{Most of the type II associated CMEs have velocities greater than 400 km/sec.}
\end{itemize}

In figure \ref{CMEShockFlare}, we present a {\emph{comprehensive}} schematic of the CME, flare and radio burst
sequence of events for the total of our data set. 

\subsection{A new Algorithm for Fine Structure Detection \& Analysis in Dymamic Spectra}

An efficient algorithm for the detection of linear and quasi--linear structures in grey-scale images was 
developed (\cite{Tsitsipis}). This algorithm introduces the 
{\em Angular Energy Density\/} and detects angular distribution of
these structures in images; this statistical approach proves to be of value in case of a great 
number of line segments. As is based on Fast Fourier Transform  methods, the computational efficiency and,
in turn, the rate of estimation are high.  

As regards solar radio burst fine structure the method proposed appears quite sensitive in slope 
detection. Due to the correspondence between frequency and height in
the Corona the slopes of bursts on the time-frequency dynamic spectra provide direct estimation
of exciter speed. This algorithm is a useful tool in the study of 
solar radio bursts wih a drift, such as type III, fibers etc; an example is demonstrated in figure \ref{FE} .

\subsection{Super Short Structures \emph{SSS}}
A new type of super--short structures (SSSs) recorded during metric type IV bursts is reported in \cite{Magdalenic}.
Their duration, at half power ranges from 4-50 ms, almost 10 times shorter than spikes at corresponding frequencies. 
The solar origin of the SSSs was confirmed by one-to-one correspondence between spectral recordings of Artemis--IV 
and high time resolution single frequency measurements (Trieste--Italy).
The recorded  of \emph{SSS} were divided in three categories:
\begin{itemize}
\item{\emph{Simple broad-band SSSs}; characterized by a bandwidth $\Delta f \geq 100~MHz$. They are further subdivided in:}
	\begin{itemize}
	\item {\emph{SSS-pulses} have duration 10--20 ms, and bandwidth $\approx~100 MHz$. 
					They appear in groups and, occasionally, exhibit quasi-periodic behaviour.}
	\item  {\emph{Drifting SSSs} have duration 30-70 ms. Their bandwidth in general exceeds the 100 MHz, and measured 
						frequency drift rates  ($|\Delta f / \Delta t| \approx$ 400-1000 MHz/sec).}
	\end{itemize}
\item{\emph{Simple narrow-band SSSs} are distinguished by their narrow bandwidth $\Delta f \le~20MHz$. They are, also, subdivided into:}
	\begin{itemize}
	\item {\emph{Spike-like SSSs} are  the shortest \emph{SSSs} with duration 4-30 ms.
		Their bandwidth is mostly $\Delta f  \le$ 20 MHz. If measurable, frequency drifts 
							are $|\Delta f / \Delta t| \ge~800~MHz/sec$.}
	\item {\emph{Patch--like SSSs} exhibit, due to their morphological diversity,
					a rather broad range of duration varying between 4 and 50 ms. The frequency
					bandwidth is $\Delta f  \le$ 15 MHz, and it can be as low as a few
					MHz. This qualifies \emph{patch--like SSS} as the \emph{SSSs} of the narrowest bandwidth. 
					Their spectral appearance varies; they can resemble dots, sails or flags and were 
					further subdivided accordingly to:  \emph{dot--like SSS},
					\emph{sail-like SSS} and \emph{flag--like SSS}.}
	\end{itemize}
\item{\emph{Complex Super Short Structures} are characterised by an emission and an absorption element. Two subcategories could be distinguished:}
	\begin{itemize}
	\item {\emph{Rain-drop bursts} consist of a narrow-band emission \emph{HEAD} ($\Delta f \approx 5 MHz$)
		and a broad-band absorption \emph{TAIL} ($\Delta f \approx 40 MHz$). The durations are approximately 50 ms for 
		the \emph{HEAD} and 30 ms for the \emph{TAIL}. Both the \emph{HEAD} and the \emph{TAIL} exhibit frequency drift 
		which is $|\Delta  f / \Delta t| \approx 60 \pm10$ MHz/sec and $|\Delta  f/\Delta t|\approx 1000
		\pm 400$ MHz/sec, respectively.}
	\item {\emph{Blinkers} are drifting bursts ($|\Delta f / \Delta t| \approx$ 650 MHz/sec)
				consisting of absorption element that is switching abruptly to emission element. 
				Opposite cases have also been found (emission in the high-frequency part, and absorption in low-frequency part of
				the burst). The duration $d_{1/2} =30-40~ms$. They are the \emph{SSSs} with the largest 
				bandwidth $\Delta f >150~MHz$.}
	\end{itemize}
\end{itemize}

\section{Conclusions \& Future Perspectives}

ARTEMIS--IV is characterized by high time and frequency resolution as well as autonomous 
operation.  We have obtained high quality data which are used in improving our 
understanding of solar radio bursts and of the underlying physics of the corona and the 
interplanetary medium. The particular areas of interest have been, transient activity and flares, 
energy dissipation, electron acceleration and transport during flares, radio signatures of CMEs
and coronal radio radiation amongst others.

The most comprehensive results have been obtained from coordinated studies  
of radio observations and data recorded in other spectral ranges. Further comparisons with  
observations of HXR, energetic electrons, EUV and SXR imaging
can put coherent emissions into context and may open exciting new possibilities for radio diagnostics
unfolding their full potential as a tool for understanding plasma processes and energy release in the solar corona.

The use of industry standard hardware and the modular architecture of the system provide a significant potential for 
further expansion. In our future plans three Acusto--Optic receivers are envisaged for full
spectral coverage.  We also intend to develop and integrate processing algorithms for supervised learning 
and pattern recognition, in an attempt to minimize recording of inactive periods and to automate
classification of observational data.


\begin{thebibliography}{}

\bibitem{Andrews}
M.~D.~Andrews,~\emph{AGU Fall Meeting}, Abstract~\textbf{SH42A--0766}, (2001).

\bibitem{Antiochos}
 S.~K.~Antiochos, C.~R.~DeVore, and~J.~A.~Klimchuk, \emph{Astrophys.J.},~\textbf{510}, 485--493,~(1999).

\bibitem{Benz}
A.~O.~Benz, \emph{Lecture Notes in Physics},~\textbf{612},~80--95, (2003)

\bibitem{Bougeret}
J.--L.~Bougeret,P.~Zarka, C.~Caroubalos et al,~\emph{GeoRL}, \textbf{25}, 2513--2516, (1998).

\bibitem{Caroubalos01a}
C.,~Caroubalos, D.,~Maroulis, N.,~Patavalis, et al, \emph{Experimental Astronomy}, \textbf{11},23--32,~(2001).

\bibitem{Caroubalos01b}
C., Caroubalos, C.~E., Alissandrakis, A. Hillaris, et al,  \emph{Sol. Phys}, \textbf{204}, 167--179, (2001).

\bibitem{Caroubalos04}
C., Caroubalos, A., Hillaris, C., Bouratzis, C.~E., Alissandrakis et al,~\emph{A\&A} \textbf{413} 1125-1133 (2004).

\bibitem{Classen}
 H.~T.,~Classen, \& H.,~Aurass,~\emph{A\&A},~\textbf{384}, 1098--1106, (2002)

\bibitem{Magdalenic}
J., Magdalenic, A., Hillaris, P., Zlobec, B., Vrsnak,~\emph{These Proceedings}

\bibitem{Pohjolainen01}
S., Pohjolainen, D., Maia, M., Pick, et al, \emph{ApJ}~\textbf{556}, 421--431,~(2001).

\bibitem{Pohjolainen05}
S. Pohjolainen, N., Vilmer, J.~I., Khan, and A.~E. Hillaris,~\emph{A\&A} \textbf{434}, 329-341,~(2005).

\bibitem{Sheeley}
N.,~Sheeley, R.~T., Stewart, R.~D.,~et al,~\emph{ApJ}, \textbf{279}, 839--847,(1984)

\bibitem{Tsitsipis}
P., Tsitsipis, A., Kontogeorgos, A., Hillaris, X., Moussas, C., Caroubalos, 
P., Preka--Papadema,~\emph{Pattern Recognition} (submitted)(2005).

\bibitem{Vrsnak}
B.,~Vrsnak, A.,~Warmuth, M.,~Temmer, A.,~Veronig, J.,~Magdalenic, A.,~Hillaris, 
M.,~Karlicky.,~\emph{A\&A}, (accepted) (2005).

\bibitem{Zhang}
J., Zhang, K.~P., Dere, R.~A., Howard, M.~R., Kundu  and S.~M., White, ~\emph{ApJ}, \textbf{559}, 452--462 (2001).

\end{thebibliography}
\end{document}